\documentstyle[12pt]{article}
\begin{document}

\makeatletter
\@addtoreset{equation}{section}
\def\theequation{\thesection.\arabic{equation}}
\makeatother

\begin{flushright}{UT-878
}
\end{flushright}
\vskip 0.5 truecm
\vskip 0.5 truecm

\begin{center}
{\large{\bf Remarks on Shannon's Statistical Inference  and 
the Second Law in
Quantum Statistical Mechanics}}
\end{center}
\vskip .5 truecm
\centerline{\bf Kazuo Fujikawa}
\vskip .4 truecm
\centerline {\it Department of Physics,University of Tokyo}
\centerline {\it Bunkyo-ku,Tokyo 113,Japan}
\vskip 0.5 truecm

\begin{abstract}
We comment on a formulation of quantum statistical mechanics, 
which incorporates the statistical inference of Shannon.
 Our basic idea is  to 
distinguish the dynamical entropy of von Neumann,
$H = -k Tr \hat{\rho}\ln\hat{\rho}$, in terms of the 
density matrix $\hat{\rho}(t)$,
and the statistical amount of uncertainty of 
Shannon, $S= -k \sum_{n}p_{n}\ln p_{n}$, 
with $p_{n}=\langle n|\hat{\rho}|n\rangle$ in the 
representation where the total energy and particle numbers are
diagonal. These quantities satisfy the inequality
$S\geq H$.
We propose to interprete Shannon's statistical 
inference as specifying the {\em initial conditions} of the 
system in terms of $p_{n}$. 
A definition of macroscopic observables which are characterized 
by intrinsic time scales is given,
and a quantum mechanical condition on the system, which ensures  
equilibrium, is discussed on the basis of time averaging. 
 An interesting analogy of the change of entroy with the 
running coupling in renormalization group is noted.  
A salient feature of our approach is 
that the distinction between 
statistical aspects and dynamical aspects of quantum statistical 
mechanics is very transparent.   
\end{abstract}

\section{Introduction}

$A$ formulation of statistical mechanics on the basis of 
Shannon's information theory ${}^{1)}$ has been proposed by 
Jaynes in 1957.${}^{2)}$
This formulation utilizes the least biased statistical 
inference about a physical system on the basis of a limited 
amount of information available. 

In the present note we discuss quantum 
statistical mechanics,${}^{3-9)}$ which
incorporates  Jaynes's proposal, and we examine what kind of 
picture appears if one
 distinguishes the  
statistical aspects and dynamical aspects of quantum statistical 
mechanics to the maximum extent. 
We work exclusively on quantum statistcal 
mechanics, though in the course of our discussion we 
comment  on the recent progress in 
the Boltzmann approach to classical statistical mechanics
.${}^{10-12)}$

We would like to briefly summarize the basic aspects of our 
analysis.   
We examine a quantum mechanical mixed state, which is slightly 
away from thermal 
equilibrium, in the representation where the total energy and 
particle numbers are diagonal. The average value of any 
macroscopic observable $\hat{O}$ is given by
$\langle \hat{O}(t)\rangle=Tr\hat{O}\hat{\rho}(t)$.
For the system slightly away from equilibrium, 
$\langle \hat{O}(t)\rangle$ is time dependent in general, and 
thus the density matrix $\hat{\rho}(t)$ is time dependent.
Following von Neumann, we introduce the (dynamical) entropy 
defined by
\begin{equation}
H=-kTr\hat{\rho}\ln\hat{\rho} 
\end{equation}
which is a generalization of Gibbs entropy to a quantum system.
It is well known that $H$ thus defined is constant in time.

The basic observation in this note is that we can define 
another quantity
\begin{equation} 
S=-k\sum_{n}p_{n}\ln p_{n}
\end{equation}
in terms of $p_{n}\equiv\langle n|\hat{\rho}|n\rangle$ 
in the 
representation where the total energy and particle numbers are 
diagonal. By definiton we have $p_{n}(t)=p_{n}(0)$ and thus 
$S$ is also constant in time. We {\em propose} to identify $S$ 
thus defined, which can be  regarded as basically statistical 
quantity, as the amount of uncertainty of Shannon which was 
introduced into statistical mechanics by Jaynes.${}^{2)}$
Note that $H$ in (1.1) agrees with $S$ in (1.2) only when 
$\hat{\rho}$ is diagonalizable simultaneously 
with the total Hamiltonian and the total particle number 
operator. This clear distinction between  $H$ in (1.1) and  $S$ 
in (1.2), to our knowledge, has not been made in the 
past.${}^{13)}$

In the present formulation of quantum statistical mechanics,
we regard the least biased estimate on the basis of limited 
amount of information discussed by Jaynes as a least biased 
estimate of {\em initial} conditions on the diagonal elements  
$p_{n}$.  
Note that the diagonal elements of $\hat{\rho}(t)$ are constant
in time and thus they cannot approach equilibrium values by
any dynamical motion but by statistical inference. (Our initial
conditions are thus final conditions also.)
The initial state is thus specified by $\{p_{n}\}$
and a set of macroscopic observables other than the total energy
and particle  number.
From the days of Boltzmann, it is well known that the second law 
of thermodynamics is mechanically represented only by means of 
assumptions regarding initial conditions,${}^{12)}$ or
by choosing
the {\em typical} states in the classical Boltzmann 
analysis.${}^{11)}$    
 In quantum statistical mechanics, as 
formulated here, it is clear that Shannon's statistical 
inference does not precede physical dynamics.

As for the main problem of quantum 
statistical mechanics as to what kind of {\em dynamical} 
properties of the 
system ensure the approach to eventual thermal equilibrium,
we propose to analyze a set of macroscopic observables, each of 
 which is characterized by an intrinsic time scale $\tau$. 
As for the entropy law, some form of coarse graining is necessary
not only in the Gibbs approach to quantum 
statistical approach${}^{8)}$ but also in the 
Boltzmann approach to classical statistical 
mechanics.${}^{10)}$ In our approach it turns out to be more
convenient to take a coarse graining in the ``time direction''
or a suitable time averaging.${}^{14,9)}$    
 The entropy law of 
Clausius in the present formulation is expressed as the 
approach of the macroscopic observables of the system on a
suitable {\em time average} to those of 
the almost equilibrium state, whose physical entropy is 
estimated by the maximum value of Shannon's $S$.

\section{Shannon's Least Biased Inference}

We consider the variable $x$ which takes the $n$ values 
$\{x_{1},x_{2},....,x_{n}\}$ and  define the probability 
$p_{i}$ for
the variable  $x$ to assume the value $x_{i}$. The 
non-negative probability $p_{i}$ is constrained by the 
condition
\begin{equation}
\sum_{i=1}^{n} p_{i} =1,
\end{equation}
which means that the total probability is unity.
We also consider a smooth  function $f(x)$ of the variable $x$, 
such as 
$f(x)=x$.

We then ask what we can say about the set of probabilities 
$\{p_{1},p_{2},,....,p_{n}\} $, if only available 
information is the 
average value  $<f>$  of  $f(x)$ defined by 
\begin{equation}
<f> \equiv \sum_{i=1}^{n} p_{i}f(x_{i}).
\end{equation}
Clearly it is impossible to determine all $p_{i}$ uniquely for a 
large 
value of $n$ since  we know only  $<f>$. 

Shannon introduced the notion of {\em amount of uncertainty} 
$S(p_{1},p_{2},....,p_{n})$ 
for the set of variables $\{p_{1},p_{2},....,p_{n}\}$, and 
he proposed to determine each 
$p_{i}$ by 
allowing  the maximum amount of uncertainty, or equivalently, 
the least bias
 for the chosen solution of $\{p_{1},p_{2},....,p_{n}\}$.
 On the basis of a composition law, Shannon 
derived the amount of uncertainty${}^{1)}$ 
( see also Appendix in ref. 2)
\begin{equation}
S(p_{1},p_{2},....,p_{n})= -k \sum_{i=1}^{n} p_{i}\ln p_{i}
\end{equation}
with a positive constant $k$. 

We now apply the above theory of inference to
statistical mechanics. Consider a closed system for which  one 
knows that the total energy is confined within a small range 
\begin{equation}
E-\frac{1}{2}\Delta E\leq E_{n}\leq E+\frac{1}{2}\Delta E
\end{equation}
with the constraint
\begin{equation}
\sum_{n}E_{n}p_{n}=E
\end{equation}
for sufficiently small $\Delta E$ (with a fixed particle number 
$N$ and a fixed volume).  
The  maximum of the Shannon's amount of uncertainty then gives 
rise to the probability for $E_{n}$
\begin{eqnarray}
&&p_{n}=\exp[-\beta E_{n}]/Z,\nonumber\\
&&Z=\sum_{E-\frac{1}{2}\Delta E\leq E_{n}\leq E
+\frac{1}{2}\Delta E}\exp[-\beta E_{n}].
\end{eqnarray} 
The parameter $\beta=\beta(E)$, which is introduced as a 
multiplier, is defined by 
\begin{equation}
E=-\frac{\partial\ln Z}{\partial\beta}.
\end{equation}
This formulation, which exhibits the temperature explicitly, is 
more convenient than the conventional formulation in 
 microcanonical ensemble.${}^{15)}$
 This formulation, 
if one assumes static
equilibrium,  is reduced to the 
microcanonical ensemble in the limit of small $\Delta E$:
The  equal {\em a priori} probabilities are obtained as 
\begin{equation}
p_{n}\simeq\frac{\exp[-\beta E]}{\Delta W(E,N)\exp[-\beta E]}
=\frac{1}{\Delta W(E,N)}
\end{equation}
and the thermodynamic relation 
\begin{eqnarray}
F&=&-\frac{1}{\beta}\ln Z\nonumber\\
&\simeq&-\frac{1}{\beta}\ln\{\Delta W(E,N)\exp[-\beta E]\} 
\nonumber\\
&=&E-\frac{1}{\beta}S/k=E-TS.
\end{eqnarray}
Here we used the Boltzmann entropy
\begin{equation}
S=k\ln\Delta W(E,N)\simeq -k \sum_{n}p_{n}\ln p_{n}
\end{equation}
where $\Delta W(E,N)$ stands for the number of quantum states 
within the above energy range.
Our determination of the time independent quantities $p_{n}$ is 
thus consistent with the microcanonical ensemble for 
sufficiently small $\Delta E$.

In the next section, we discuss a formulation of quantum
statistical mechanics which incorporates the above statistical
inference, and yet a logically consistent 
formulation in the sense that statistical inference does not
precede dynamical time development. Note that we rely on the 
statistical inference since we deal with an enormous number of 
states $E_{n}$ in (2.4). 

\section{Statistical Inference and Quantum 
Statistical Mechanics}
 
We discuss how to describe {\em near-equilibrium} states and their
approach to thermal equilibrium in a framework 
which incorporates the least biased statistical inference.
We assume that we analyze a physical system which is completely
characterized by its total energy and particle number, if the 
thermal equilibrium should be realized for a fixed volume. 
In the representation where the total energy and the particle 
number are
 diagonal, we have the density matrix${}^{16)}$
 $\hat{\rho}(t)$ which 
satisfies${}^{8)}$ 
\begin{eqnarray}
&&Tr\hat{\rho}(t)=\sum_{n}\hat{\rho}(t)_{nn}=\sum_{n}p_{n}=1,
\nonumber\\
&&Tr\hat{\rho}\hat{{\cal H}}=\sum_{n}E_{n}p_{n}=\langle E\rangle,
\nonumber\\
&&Tr\hat{\rho}\hat{\nu}=\sum_{n}\nu_{n}p_{n}=\langle N\rangle.
\end{eqnarray}
We here assume for simplicity the presence of only one kind of 
particles.
We first note that $p_{n}$ is time independent
$p_{n} = \langle n|\hat{\rho}(t)|n\rangle=\langle n|
e^{-i\hat{{\cal H}}t/\hbar}
\hat{\rho}(0)e^{i\hat{{\cal H}}t/\hbar}|n\rangle
= \langle n|\hat{\rho}(0)|n\rangle$, 
since the total Hamiltonian is diagonal in the present
representation. We assume that either none of the energy levels 
are degenerate, or if some of them are degenerate, the density
matrix $\hat{\rho}(t)$ is diagonalized by a (constant) unitary 
transformation beforehand in each sector 
which contains the degenerate energy levels. Consequently, all 
the possible off-diagonal elements of the density matrix 
$\hat{\rho}(t)$ are time dependent.
Our {\em proposal} is to define the Shannon's amount 
of uncertainty, which is based on the information available, by
\begin{equation}
S=-k\sum_{n}p_{n}\ln p_{n}.
\end{equation}
This $S$, which carries no characteristic properties of quantum 
theory, may be assigned a purely statistical meaning, and it is 
time independent.

The dynamical entropy of von Neumann 
\begin{equation}
H=-kTr\hat{\rho}\ln\hat{\rho}
\end{equation}
which is a quantum generalization of Gibbs entropy, is also time 
independent since the time development of 
$\hat{\rho}$ is a unitary transformation. In ref. 9 
(and also in ref. 5), 
this entropy $H$, which in principle contains the effects of 
quantum coherence, is called the information entropy. In this 
paper we stick to the classical notion of information and thus to 
the amount of uncertainty defined in (3.2). The advantage of 
choosing (3.2) as the amount of uncertainty becomes clear later.
The average value of 
any operator in the Schr\"{o}dinger representation is defined by
\begin{equation}
\langle \hat{O}(t)\rangle = Tr\hat{{\rho}}(t)\hat{O}(0).
\end{equation}

In the framework of Shannon's least biased statistical inference 
used by Jaynes, the maximum value of the amount of 
uncertainty, which is identified with  $S$ in (3.2) in the 
present formulation, is considered with the constraints (3.1). 
For a {\em closed system} we are considering, the constraints are 
actually replaced by the conditions (2.4) and (2.5) in Section 2. 
One then obtains the standard result (2.6) for the constant 
diagonal elements of $\hat{{\rho}}(t)$
\begin{eqnarray}
&&p_{n}=\exp[-\beta(E)E_{n}]/Z,\nonumber\\
&&Z=\sum_{E-\frac{1}{2}\Delta E\leq E_{n}\leq E
+\frac{1}{2}\Delta E}\exp[-\beta(E) E_{n}]
\end{eqnarray}
but the time dependent off-diagonal elements of 
$\hat{{\rho}}(t)$ are left completely unspecified, namely, we 
remain maximally noncommittal with regard to missing 
information.${}^{17)}$

On the other hand, the maximum of the von Neumann 
entropy $H$  with conditions (2.4) and (2.5) would give rise 
to${}^{6,7,8)}$ 
\begin{eqnarray}
\langle n|\hat{\rho}(t)|n\rangle&=&\exp[-\beta(E)E_{n}]/Z,
\nonumber\\
\langle n|\hat{\rho}(t)|m\rangle&=&0,\ \ \ \ n\neq m.
\end{eqnarray}
This density matrix $\hat{{\rho}}$ is completely diagonal; in 
other words, if one imposes the 
maximum condition on the von Neumann entropy, we arrive 
at the conventional microcanonical ensemble without any freedom 
of time development. In fact, the 
conventional analysis of statistical mechanics utilizes this 
property of the entropy $H$ and attempts to prove the Boltzmann's 
H-theorem for (a coarse grained form of) $H$ in (3.3) as an 
indicator of the general tendency toward 
thermal equilibrium.${}^{8,9,18)}$
In contrast, we here attempt to characterize the approach to 
 thermal equilibrium from a different perspective. Also, we will 
later suggest that the {\em dynamical} density matrix does not 
approach the static form (3.6) even in thermal equilibrium, since
the oscillations with microscopic time scales, which should be 
described by the dynamical density matrix, always exist in the 
system.  
  
For the general situation, the maximum uncertainty inference of 
Shannon as formulated here does not specify the density matrix 
completely, and the best estimate of the average of a general 
macroscopic operator $\hat{O}$ is
\begin{equation}
\langle \hat{O}\rangle_{0}\equiv \sum_{n}p_{n}\hat{O}(0)_{nn}
=\sum_{n}\hat{O}(0)_{nn}\exp[-\beta(E) E_{n}]/Z
\end{equation}
where $p_{n}$ is defined in (3.5) and the sum is taken over 
the available eigenstates in the 
representation where the total Hamiltonian and particle number 
are diagonal. This quantity is time independent by definition and
 agrees with the {\em conventional} average in thermal 
equilibrium. 
We will 
later show that, only after a suitable time averaging, the true 
average (3.4) is well approximated by the conventional thermal 
average (3.7) if the system satisfies  certain dynamical 
properties.   
It is clear that the least biased inference of Shannon, as 
formulated here, is purely statistical and does not provide any 
information about dynamical time development.

We next note an inequality between $S$ and $H$ (see also 
refs. 5 and 19)
\begin{equation}
H\leq S.
\end{equation}
This relation is shown by using the standard technique
of the statistical mechanics${}^{20)}$.
This inequality, $H\leq S$, valid for {\em any} 
$\hat{\rho}(t)$ 
suggests that we can impose the maximum amount of uncertainty 
condition on $S$ without any dynamical constraint on $H$.
 (In contrast, if one should use (3.3) 
as Shannon's amount of uncertainty, the statistical inference 
would precede physical dynamics, since the statistical inference
would then determine the time dependent off-diagonal elements
of $\hat{\rho}(t)$ as well.) Besides, $H\leq S$ shows that we 
start with an initial state with  smaller entropy.${}^{12)}$
\\

In analogy with the definition of a quantum state in terms of a 
complete set of commuting hermitian operators,${}^{22)}$ we 
assume that our density matrix $\hat{\rho}(t)$, which is a 
generalization of the Schr\"{o}dinger wave function, is well 
specified by a set of {\em macroscopic} observables $\{\hat{O}\}$
, and total energy and particle number; in the present case, the 
operators $\{\hat{O}\}$ do not commute with the total Hamiltonian
 by our assumption. 
As in  classical Boltzmann statistical mechanics,${}^{11)}$
 where one works exclusively on macroscopic
 variables,
one may  define macroscopic observables $\hat{O}_{\tau}$ in 
the present framework by the condition
\begin{equation}
Tr\hat{\rho}(t)\hat{O}_{\tau}=
\frac{1}{\tau}\int_{t}^{t+\tau}dtTr\hat{\rho}(t)\hat{O}_{\tau}.
\end{equation}
This condition is written in full detail as
\begin{eqnarray}
\langle\hat{O}_{\tau}(t)\rangle&=&\sum_{m,n}\rho_{mn}(0)
\langle n|\hat{O}_{\tau}|m\rangle\exp[i(E_{n}-E_{m})t/\hbar]
\nonumber\\
&=&\frac{1}{\tau}\int_{t}^{t+\tau}dt\sum_{m,n}\rho_{mn}(0)
\langle n|\hat{O}_{\tau}|m\rangle\exp[i(E_{n}-E_{m})t/\hbar]
\end{eqnarray}
and thus the macroscopic observables are not sensitive to the 
microscopic time (shorter than $\tau$) dependence of 
the density matrix.${}^{23)}$
Each macroscopic observable in our definition is labeled by a 
characteristic time
scale $\tau$, which may in general depend on the temperature 
contained in the diagonal components $\{p_{n}\}$ of the density
matrix.${}^{24)}$ 
An operator with large $\tau$ gives a macroscopic
observable which agrees with our intuitive understanding: For 
example, $\tau=\infty$ for the total energy or particle number. 

We then define the non-equilibrium state 
{\em operationally} by the relation
\begin{eqnarray}
\langle \Delta\hat{O}_{\tau}(t)\rangle&\equiv& Tr\hat{\rho}(t)
[\hat{O}_{\tau}-
\sum_{n}p_{n}<n|\hat{O}_{\tau}|n>]\nonumber\\
&=&Tr\hat{\rho}(t)\hat{O}_{\tau}-
\sum_{n}p_{n}<n|\hat{O}_{\tau}|n> \neq 0
\end{eqnarray}
for some macroscopic observables $\hat{O}_{\tau}$ other than the 
total 
energy and particle number. Here $p_{n}$ is defined in (3.5). 
If we do not find any sensible macroscopic observable 
$\hat{O}_{\tau}$ 
which satisfies the above relation ( after a suitable time 
averaging described later ), the system is in thermal 
equilibrium.

Our system described by the density matrix $\hat{\rho}(t)$ then
develops with time following  Schr\"{o}dinger equation with a 
fixed value of the von Neumann entropy $H$.

\section{Second Law in Quantum Statistical Mechanics}

Our next task is to {\em specify} what kind of dynamical 
properties of a many particle system ensure that the system 
with initial conditions defined by our statistical inference will
 in the long run approach the almost equilibrium state. 
We first note that the time average of our $\hat{\rho}(t)$ 
over a sufficiently long period approaches arbitrarily close 
 to the 
equilibrium $\hat{\rho}_{0}$ with diagonal elements $p_{n}$ in 
(3.5). In this sense, our system by its construction satisfies 
the Boltzmann's ergodic postulate; the time average 
behavior of a system is the same as its equilibrium behavior. 
Namely a suitable time averaging  of (3.11) gives rise to 
\begin{equation}
\langle\overline{\Delta\hat{O}_{\tau}}\rangle\simeq 0
\end{equation}
which is the (necessary) condition for equilibrium.

We need to sharpen  the time averaged behavior (4.1) to be 
a sufficient dynamical condition for equilibrium.  
We first recall a  quantum version of Poincare's recurrence 
theorem.${}^{25)}$ The theorem states that observables in
 a (finite) many particle system with discrete energy spectrum 
are
almost periodic, namely, after a suitable time lapse the system 
comes back to arbitrarily close to the original configuration.
This means that the system has a rather well defined dynamical 
property and we here deal with those finite systems.${}^{26)}$
 We however assume that 
the recurrence time for a many particle system is sufficiently 
long by the time scale of our laboratory. The reccurence theorem
provides a partial justification for the independent specification
of diagonal elements of $\hat{\rho}$ in (3.5) and the 
non-diagonal elements in (3.11), since an apparently thermal
equilibrium state, which is related to (3.5), can come back to 
any original starting configuration. 

To ensure the thermal equilibrium, we need to avoid  the 
persistent synchronized collective oscillation even if the 
averaged behavior (4.1) is satisfied. We expect that   
the probability for a great number of oscillators in 
$\hat{\rho}(t)$ to synchronize persistently is  
negligibly small for a system of a many particle system.
 In the generic 
situation, our system  is 
expected to give  a negligible time correlation between 
$\langle \Delta\hat{O}_{\tau}(t)\rangle$ for a large time 
difference $|t_{1}-t_{2}|$, which is however very small compared 
to  the recurrence time. 

The averaged behavior  (4.1) is now sharpened to be a stronger 
{\em dynamical postulate} as 
( see also (3.11))  
\begin{eqnarray}
|\langle\overline{\Delta\hat{O}_{\tau}}\rangle(t,\Delta t_{c})|
&\equiv&|\frac{1}{\Delta t_{c}}
\int_{t}^{t+\Delta t_{c}}dt \langle \Delta\hat{O}_{\tau}(t)
\rangle|
\nonumber\\
&=&|\frac{1}{\Delta t_{c}}
\int_{t}^{t+\Delta t_{c}}dtTr\hat{\rho}(t)\hat{O}_{\tau}-
\sum_{n}p_{n}<n|\hat{O}_{\tau}|n>|\nonumber\\
&\ll& |\langle \hat{O}_{\tau}\rangle_{0}|
\end{eqnarray}
for {\em any} macroscopic observable 
$\hat{O}_{\tau}$ and a fixed finite $\Delta t_{c}$;
$\langle \hat{O}_{\tau}\rangle_{0}
=\sum_{n}p_{n}<n|\hat{O}_{\tau}|n>|\nonumber$  is defined in 
(3.7) with $p_{n}$ in (3.5). For
$\Delta t_{c}=\infty$, the left-hand side of this relation 
vanishes by our construction. We impose this condition for a 
{\em finite} $\Delta t_{c}$ and assume that 
$\langle\overline{\Delta\hat{O}_{\tau}}\rangle(t,\Delta t_{c})$ 
is not sensitive to the absolute value of $t$ and a small 
variation of $\Delta t_{c}$. 
Physically, this condition means the existence of a 
{\em well-defined relaxation time} for a set of macroscopic 
observables.  
The actual magnitude of $\Delta t_{c}$, which is expected to be 
microscopically quite long and of the order of macroscopic time
scale, will generally depend on the 
specific system we are analyzing. 
In terms of the parameter $\tau$ in (3.9), we expect (for an
operator with a finite $\tau$)
\begin{equation}
\tau< \Delta t_{c}<\infty.
\end{equation}
To observe the relaxation in terms of the macroscopic observables
 $\hat{O}_{\tau}$, we need to have $\tau< \Delta t_{c}$. 

Note that
the condition (4.2) does not contradict the reccurence 
theorem:${}^{25)}$
In practice, after the initial relaxation, it might be that one 
can take $\Delta t_{c}\sim\tau$, namely, one cannot recognize 
the sizable deviation from thermal equilibrium by the time 
resolution of macroscopic observables. However, if the 
reccurence occurs, one 
need to wait for the time $\sim\Delta t_{c}$ for the system 
to relax again.

Since
\begin{equation}
\langle\hat{O}_{\tau}(t)\rangle=\sum_{m,n}\rho_{mn}(0)
\langle n|\hat{O}_{\tau}|m\rangle\exp[i(E_{n}-E_{m})t/\hbar]
\end{equation}
we have only the near diagonal components after the above time
averaging (4.2); namely, only the terms with 
\begin{equation}
|E_{n}-E_{m}|\leq 2\pi\hbar/\Delta t_{c}
\end{equation}
survive the time averaging. Note that $2\pi\hbar/\Delta t_{c}$ is
very small compared to $\Delta E$ in (3.5) for a many
particle system we are interested in.${}^{27)}$
Also the real 
time-independent diagonal components
are subtracted in $\langle \Delta\hat{O}_{\tau}(t)\rangle$. In 
the remaining terms we have a sum of 
complex amplitudes 
$\rho_{mn}(0)\langle n|\hat{O}_{\tau}|m\rangle$ with 
nearly equal frequencies $(E_{n}-E_{m})/\hbar$: Our 
condition (4.2) is that the sum of oscillating 
quantities are either absent or destructively 
interfere
\begin{equation}
|\sum_{0<|E_{n}-E_{m}|\leq 2\pi\hbar/\Delta t_{c}}\rho_{mn}(0)
\langle n|\hat{O}_{\tau}|m\rangle\exp[i(E_{n}-E_{m})t/\hbar]|
\ll |\langle \hat{O}_{\tau}\rangle_{0}|
\end{equation}
for a general value of $t$.

As a concrete example of our analysis, we illustrate the problem 
of a gas confined 
in the left-half of a box and then removing the partition, 
although this probelm corresponds to the case of far from 
equilibrium.${}^{28)}$
We make a statistical inference to fix $p_{n}$ on the basis of 
the information 
about all the possible energy spectra of the {\em entire} box, 
the average energy and particle number. The macroscopic 
observable $\hat{O}_{\tau}(\vec{x})$ may be chosen as the 
particle number density (in a suitably smeared sense) inside the 
entire box. We define 
$\langle \hat{O}_{\tau}\rangle_{0}(\vec{x})$ by using the result 
of the above inference and (3.7). The observable 
\begin{equation}
Tr\hat{\rho}(t)\Delta\hat{O}_{\tau}(\vec{x})=Tr\hat{\rho}(t)
\hat{O}_{\tau}(\vec{x})
-\langle \hat{O}_{\tau}\rangle_{0}(\vec{x}),
\end{equation}
which has a positive peak in the left-half of the box 
and a negative peak in the right-half of the box at $t=0$, is 
described by choosing the off-diagonal time dependent elements 
of $\hat{\rho}(t)$ suitably. Our inference agrees with
 the conventional answer of statistical mechanics, if the 
possible macroscopic oscillation in 
$Tr\hat{\rho}(t)\Delta\hat{O}_{\tau}(\vec{x})$ diminishes soon. 

To analyze the effects of time averaging,
one may consider a simpler example of the average position of 
particles (instead of the particle number density)
\begin{equation}
\hat{O}=\vec{X}=\sum_{i}\vec{x}_{i}/N
\end{equation}
of the gas confined into the left-half of the box at $t=0$, 
which is one 
of the indicators of the macroscopic motion of particles. One may
make a crude estimate of the typical frequency contained in 
$\vec{x}_{i}$
by considering the matrix element 
$\langle n+1|\vec{x}_{i}|n\rangle$  for a free particle as 
\begin{equation}
\omega\sim (\hbar\pi(n+1))^{2} /(2mL^{2}\hbar)-
(\hbar\pi n)^{2} /(2mL^{2}\hbar)\sim (\hbar\pi n)/(mL^{2})
\sim v/L
\end{equation}
where $v=(\hbar\pi n)/mL$ is the {\em typical} velocity, which 
is 
determined by the temperature appearing in the diagonal elements 
$p_{n}$. Here $L$ is the size  of the box. It is then unlikely
that $\langle\hat{O}(t)\rangle$ contains the sizable components 
with frequency 
\begin{equation}  
\omega=|E_{n}-E_{m}|/\hbar\leq 2\pi/\Delta t_{c}
\end{equation}
to survive the time averaging for $\Delta t_{c}\gg L/v$ (or
equivalently, $2\pi/\Delta t_{c}\ll 2\pi v/L$), 
except for the static diagonal elements.  For this choice of 
$\hat{O}$ and the crude 
estimate, the gas which was in thermal equilibrium and confined 
into the left-half of the box at $t=0$ satisfies the condition 
(4.2) for $\Delta t_{c}\gg L/v$. The relaxation time in the 
present example is determined by the typical transport time scale 
$\tau\sim L/v$, provided that a soliton-like persistent collective
motion does not occur.

\subsection{Entropy law of Clausius}

The physical entropy of the final thermodynamic state defined by 
this time averaging (4.2), which is characterized by 
$\Delta t_{c}$, is estimated by 
\begin{equation}
\bar{H}(\Delta t_{c})\equiv -k Tr \bar{\hat{\rho}}
\ln \bar{\hat{\rho}}
\end{equation}
with 
\begin{equation}
\bar{\hat{\rho}}=(1/\Delta t_{c})
\int_{t}^{t+\Delta t_{c}}\hat{\rho}(t)dt
\end{equation}
for a generic value of $t$. This value is expected 
to be close to the maximum of Shannon's 
statistical amount of uncertainty $S$, 
which is the maximum value of any sensible definition of entropy 
because 
of (3.8): Note that 
\begin{equation}
S\geq \bar{H}(\Delta t_{c})
\end{equation}
since 
$\bar{\hat{\rho}}_{nn}=\hat{\rho}(t)_{nn}=p_{n}$ and 
$Tr\bar{\hat{\rho}}=1$, which are sufficient to prove (3.8). We 
interprete 
this approach of $\bar{H}(\Delta t_{c})$ to $S$ as a 
manifestation of the entropy 
law in the present formulation of quantum statistical 
mechanics.${}^{29)}$
We reiterate that we used two ingredients to formulate the entropy
law in our approach : The first is the statistical input 
related to the least biased inference on the basis of a limited
amount of information available (3.5), and the second is the 
dynamical input related to the time averaging in (3.9) and (4.2).

The von Neumann entropy $H$ is in contrast  rigidly defined by 
the basic dynamics, and it does not 
allow any arbitrary manipulation such as taking a time averaging
of $\hat{\rho}(t)$. The dynamical entropy $H$ of 
von Neumann stays constant throughout the unitary time 
development of the system regardless of our time averaging
procedure. In fact, the above time averaging (4.2) to define the 
physical thermodynamic state resolves the 
discrepancy of the dynamical von Neumann $H$ and the physical
statistical entropy (4.11) we defined. 
Physically, the von
Neumann entropy $H$ is sensitive to the dynamical motion of all
the time scales in the system, whereas the thermodynamic entropy
(4.11) is not sensitive to the motion with time scales 
shorter than $\sim \Delta t_{c}$; moreover, the condition (4.2) 
states that the macroscopic motion with time scales larger than 
$\sim \Delta t_{c}$ in the system is negligible. 

 We here note an interesting analogy of the
present formulation of physical entropy with
the renormalization group in field theory. The parameter 
$\Delta t_{c}$ (or to be precise, $\hbar/\Delta t_{c}$) 
characterizes the energy scale of the theory, and 
the entropy $\bar{H}(\Delta t_{c})\equiv -k Tr \bar{\hat{\rho}}
\ln \bar{\hat{\rho}}$ in (4.11) corresponds to the renormalized 
running
coupling constant. The ultra-violet limit 
$\Delta t_{c}\rightarrow 0$ gives rise to the von Neumann 
entropy $H$, which corresponds to the bare coupling constant, 
the fundamental quantity defined by the basic dynamics , namely, 
quantum mechanics in the present case. But physics is not 
sensitive to the bare coupling constant. The infrared limit of 
$\bar{H}$ for $\Delta t_{c}\rightarrow large$ gives rise to the 
measurable quantity, the maximum of Shannon's $S$, corresponding 
to the coupling constant $\alpha=1/137$ in QED defined in the 
Thomson limit. In this analogy, the entropy law of Clausius 
corresponds to a statement of the existence of a stable 
infrared fixed point for finite $\Delta t_{c}$. 

This picture also suggests that 
the dynamical density matrix $\hat{\rho}(t)$ does not approach
the static diagonal form in (3.6) even in thermal equilibrium, 
since the oscillations with microscopic time scales always exist.
The microscopic time dependence of the {\em equilibrium} density 
matrix is also expected in the conventional Gibbs ensemble
theory, if one defines the thermal equilibrium 
by macroscopic observables $\hat{O}_{\tau}$ with well defined
time scale $\tau$. Our definition of the macroscopic observables 
$\hat{O}_{\tau}$ in (3.10) shows that
 the macroscopic observables are not sensitive to the 
possible microscopic time (shorter than $\tau$) dependence of 
the equilibrium density matrix.

Our picture is expected to have an  implication on the 
linear response theory for time dependent current 
correlations${}^{30)}$ with frequencies which are comparable 
to $\sim 1/\tau$, since the possible time dependence of the 
equilibrium density matrix is not ignored for such a case.
In the applications of the fluctuation-dissipation theorem,
 one usually examines the correlation functions of operators 
averaged with the static diagonal equilibrium density 
matrix $e^{-\beta\hat{{\cal H}}}/Z$. The possible 
microscopic time 
dependence of equilibrium density matrix may lead to interesting 
implications on the analysis of the foundations of linear
response theory.

\section{Discussion}

The purpose of this note has been to discuss quantum
statistical mechanics which incorporates Shannon's statistical
inference, and to analyze a resulting 
picture for the second law. Our statistical inference 
does not resolve the basic issue why the statistical inference 
works for a many particle system, but our statistical inference 
,unlike the equal {\em a priori} probabilities, allows 
an analysis of the dynamical aspects of the second law.

Our picture for the general tendency toward thermal equilibrium 
 is quite different from the one in the conventional formulation
 of quantum mechanical 
H-theorem:${}^{8,9,18)}$ 
 We note that the conventional coarse-grained approach 
 performs not only the statistical operation by 
setting the diagonal elements in each subsector of the density 
matrix to be equal by assuming equal {\em a priori} 
probabilities, but also the dynamical operation by setting 
the off-diagonal {\em time dependent} elements to be zero
to let the entropy increase. 
  
We now briefly comment on the two ambitious approaches to the 
second
law on the basis of purely dynamical considerations, to be 
compared  with our more conservative statistical analysis. The 
first is the approach of Tasaki${}^{31)}$ and the second 
is the approach of Van Hove:${}^{32)}$

Tasaki analyzed a possible quantum mechanical derivation of 
a canonical ensemble starting from a pure quantum state.
The basic idea is to consider two quantum systems, a bath 
$|\alpha\rangle$ and a subsystem $|a\rangle$, and one then 
introduces a suitable interaction between
them so that one obtains a pure quantum state for the combined 
system,
\begin{equation}
|\psi\rangle=\sum_{a,\alpha}M_{\alpha,a}
|\alpha\rangle\otimes|a\rangle
\end{equation}
which has a vanishing von Neumann entropy. One then 
traces out the bath system in the density matrix
\begin{equation}
\hat{\rho}_{tot}=|\psi\rangle\langle\psi|\rightarrow
\hat{\rho}_{sub}=\sum_{a,b}(M^{\dagger}M)_{ab}|b\rangle\langle a|.
\end{equation}
Because of the quantum entanglement of two systems, one then 
obtains a mixed state for the subsystem which has a non-vanishing
entanglement entropy${}^{33)}$
\begin{equation}
H_{sub}=-kTr\hat{\rho}_{sub}\ln\hat{\rho}_{sub}>0.
\end{equation}
To reproduce a canonical ensemble for the 
subsystem, Tasaki assumes the ``hypothesis of 
equal weights for eigenstates'' for the combined system. He also
argues the cancellation of oscillating non-diagonal components 
of the density matrix for the subsystem at sufficiently large 
$t$ by using Chebyshev's inequality.${}^{31)}$
If one can show that the  ``hypothesis of equal weights for 
eigenstates'' holds for a rather general class of dynamical 
systems, the quantum entanglement entropy would provide a 
physical explanation of the statistical entropy.
At this moment, it appears
that the generality of the hypothesis has not been established.   

In the analysis of entanglement entropy, a clear distinction 
between the von Neumann's $H$, which is equal for both of the 
bath and the subsystem${}^{33)}$ $H_{bath}=H_{sub}$ and 
thus not extensive, and the Shannon's $S$, 
which could be vastly different for the bath and the subsystem
and thus could be extensive, is expected to be essential.

Van Hove${}^{32)}$ analyzed the possible approach of a 
general mixed 
state to microcanonical states by a dynamical time development
in the limit $t\rightarrow\infty$. To be specific, he starts
with a mixed state
\begin{equation}
\hat{\rho}(0)\equiv \int d\alpha |\alpha\rangle |c_{\alpha}|^{2}
\langle\alpha|
\end{equation}
where $\hat{H}_{0}|\alpha\rangle=\epsilon_{\alpha}|\alpha\rangle$
, and the unitary time development generated by 
\begin{equation}
U(t)=\exp [-i(\hat{H}_{0}+\lambda\hat{V})t/\hbar].
\end{equation}
His major claim is ( see Eq. (1.4) in the first paper 
in ref. 32) 
\begin{equation}
Tr\hat{O}U(t)\hat{\rho}(0)U^{-1}(t)\rightarrow 
\langle \hat{O}\rangle_{microcanonical}
\end{equation}
for $t\rightarrow\infty$. Since the operator $U(t)$ is unitary
for whatever large but finite $t$, both of the von Neumann's $H$ 
and Shannon's $S$ remain at the 
initial value different from the microcanonical value.
Van Hove however considers a {\em singular} limit 
$\lambda^{2}t\rightarrow\ \ constant$ for $t\rightarrow\infty$ 
(i.e., $\lambda t\sim 1/\lambda\rightarrow\infty$ for 
$t\rightarrow\infty$).
In such a  limit, if literally taken, the $S$-matrix
\begin{equation}
\hat{S}=\lim_{t_{\pm}\rightarrow\pm\infty}
e^{i\hat{H}_{0}t_{+}/\hbar}
e^{-i(\hat{H}_{0}+\lambda \hat{V})(t_{+}-t_{-})/\hbar}
e^{-i\hat{H}_{0}t_{-}/\hbar}
\end{equation}
(and consequently perturbation theory he uses) is not defined,
since the condition of an adiabatic
switch-on and switch-off of the interaction is not satisfied. 
See also ref. 34.
If one can provide a mathematical basis for the singular limit,
one would be able to derive the microcanonical ensemble
from a general mixed state by a unitary time development. At this
moment, to our knowledge, such a mathematical basis appears to be
 missing.

Finally, we mention recent activities on an alternative approach 
to the second law. Jarzynski${}^{35)}$ found the following 
amusing identity
\begin{eqnarray}
\frac{Z_{1}}{Z_{0}}&=&\frac{1}{Z_{0}}\int d\mu(z^{\prime})
e^{-\beta H_{1}(z^{\prime}) }\nonumber\\
&=&\frac{1}{Z_{0}}\int d\mu(z^{\prime})
e^{-\beta H_{0}(z)-\beta ( H_{1}(z^{\prime})-H_{0}(z))}\nonumber\\
&=&\frac{1}{Z_{0}}\int d\mu(z)
e^{-\beta H_{0}(z)-\beta W(z)}=\langle e^{-\beta W(z)}\rangle 
\end{eqnarray}
where we defined the mapping
\begin{equation}
z=(q(t_{0}),p(t_{0}))\rightarrow 
z^{\prime}=(q^{\prime}(t_{1}), p^{\prime}(t_{1}))
\end{equation}
as the canonical transformation
generated by the time dependent Hamiltonian (a quantum version 
is also known${}^{36)}$)
\begin{equation}
H_{\lambda(t)}, \ \ \ t_{0}\leq t \leq t_{1}
\end{equation}
with
\begin{equation}
\lambda(t_{0})=0, \ \ \ \lambda(t_{1})=1.
\end{equation}
The Liouville theorem $d\mu(z^{\prime})=d\mu(z)$ is essential 
in the above identity.
We also defined the work done during the time development by
$W(z)\equiv  H_{1}(z^{\prime})-H_{0}(z)$.

If one defines the Helmholtz free energy for the system described
 by Hamiltonian $H_{1}$ at temperature $\beta$ by $F_{1}(\beta)$
, one obtains from (5.8)
\begin{equation}
\Delta F(\beta)=F_{1}(\beta)-F_{0}(\beta)
=-\frac{1}{\beta}\ln \langle e^{-\beta W(z)}\rangle.
\end{equation}
and, by noting the mathematical inequality 
$\langle \exp[A(z)]\rangle\geq 
\exp[ \langle A(z)\rangle]$,
\begin{equation}
\Delta F(\beta)\leq \langle W \rangle
\end{equation}
which resembles the basic thermodynamic inequality,${}^{6)}$
 an alternative expression of 
the second law.
The identity (5.8) as it stands is equivalent to the Liouville
theorem and thus contains no information about the thermal
entropy generation. In fact, it is known (from an explicit 
analysis
of harmonic oscillators, for example) that the equality sign in 
(5.13) does
not hold for an infinitely slow adiabatic 
work${}^{35)}$, which is by itself consistent: But to analyze 
the equality sign in (5.13), one need to analyze 
 the approach of a system in (5.8), once driven out of 
equilibrium by an external work, to thermal equilibrium again. 
The analysis of Jarzynski is thus complementary to the analysis 
in the present note, and certainly it does not replace our 
analysis. See recent works${}^{37,38}$ 
related to the above identity. See also Lenard${}^{39)}$ for an
analysis of similar inequality associated with a canonical 
ensemble.${}^{40)}$

In conclusion, we have illustrated a physical picture of 
the second law when one makes a clear distinction between 
statistical aspects and dynamical aspects in statistical 
mechanics, which is made 
possible if one uses $S$ in (3.2).
Our basic view as presented here is rather conservative, namely, 
it is based on the premise that the entropy law of Clausius is 
not a direct consequence of  microscopic dynamical laws alone. 
In the context of classical Boltzmann approach, this view appears
 to be shared with experts just to quote `` It 
follows that the macroscopic dynamics 
{\em cannot} be a consequence of the microscopic dynamics 
alone''.${}^{10)}$ 
 The remaining basic issue in the present approach is 
to specify precisely the class of many-particle Hamiltonians 
which ensure (4.2).\\ 

I thank H. Tasaki for numerous clarifying comments, and 
A. Shimizu and M. Ueda for helpful comments at the initial stage
of this work.

\appendix


\begin{thebibliography}{99}
\bibitem{f1}
There appears an important difference between 
$H$ and $S$ if one allows a time dependent Hamiltonian. The 
density matrix is still transformed by a unitary 
transformation and thus $H$ stays constant. But $S$ generally
changes except for an infinitely slow adiabatic variation of 
the Hamiltonian.
\bibitem{f2}
One can confirm the relation
$\frac{dS}{dE}=k\beta(E)$ for this solution, and we assume 
 $\beta(E)>0$.
\bibitem{f3}
One may start with  normalized pure states in the energy 
representation
$|\Psi_{k}\rangle=\sum_{n}a_{k,n}|n\rangle$, and one may define 
a density matrix for a mixed state by 
$\hat{\rho}=\sum_{k}b_{k}|\Psi_{k}\rangle\langle\Psi_{k}|
=\sum_{k}\sum_{m}\sum_{n}b_{k}a_{k,m}a^{\star}_{k,n}
|m\rangle\langle n|\equiv\sum_{m,n}\rho_{m,n}|m\rangle
\langle n|$ with $b_{k}>0$ and $\sum_{k}b_{k}=1$. It is 
however common to 
define a mixed state by using more general $\rho_{m,n}$, which
 we adopt in the present note.
\bibitem{f4}
This statistical inference of diagonal 
elements means 
that the macroscopic observables $\langle E\rangle$ and 
$\langle N\rangle$ are insensitive to whether one uses the 
original density matrix or the density matrix after one applied 
the statistical inference.
\bibitem{f5}
A sketch of the proof is as follows: Define $f(x)=-kx\ln x$ for 
$0\leq x\leq 1$, which is convex in the sense 
$f^{\prime\prime}(x)<0$ for $0< x\leq 1$. Define any 
orthonormal set of states $\{|n\rangle \}$  in terms of the 
normalized eigenstates $\varphi_{k}$ of $\hat{\rho}$ as,  
$|n\rangle=\sum_{k}a(n)_{k}\varphi_{k}, \ \ 
\sum_{k}a(n)_{k}a(l)^{\star}_{k}=\delta_{nl},\ \
\hat{\rho}\varphi_{k}=\lambda_{k}\varphi_{k}, \ \ 
\lambda_{k}\geq 0, \ \ \sum_{k}\lambda_{k}=1.$
If one defines a tangential line to $f(x)$ at 
$x=\langle n|\hat{\rho}|n\rangle
=\sum_{k}|a(n)_{k}|^{2}\lambda_{k}$ 
by $\bar{f}(x)=ax+b$, one has
$f(\lambda_{k})\leq\bar{f}(\lambda_{k})$ and 
$\langle n|\bar{f}(\hat{\rho})|n\rangle
=\sum_{k}|a(n)_{k}|^{2}\bar{f}(\lambda_{k})
=\bar{f}(\langle n|\hat{\rho}|n\rangle)
=f(\langle n|\hat{\rho}|n\rangle)$.
One can then prove${}^{21)}$
$\langle n|f(\hat{\rho})|n\rangle=\sum_{k}|a(n)_{k}|^{2}
f(\lambda_{k})\leq \bar{f}(\langle n|\hat{\rho}|n\rangle)
=f(\langle n|\hat{\rho}|n\rangle)$,
where the equality sign 
holds only for $ \hat{\rho}|n\rangle=\lambda_{n}|n\rangle $.

One thus concludes $\sum_{n}\langle n|f(\hat{\rho})|n\rangle
=-kTr\hat{\rho}\ln\hat{\rho}\leq 
\sum_{n}f(\langle n|\hat{\rho}|n\rangle)
=-k\sum_{n}\langle n|\hat{\rho}|n\rangle
\ln\langle n|\hat{\rho}|n\rangle$, as required.
\bibitem{f6}
Intuitively, one can see a possible
origin of the violation of time reversal symmetry in this time 
averaging. One cannot reproduce the original density matrix form 
the time averaged one, and in this sense the use of macroscopic
observables induces the violation of microscopic time reversal 
invariance.
\bibitem{f7}
To be more precise, the intrinsic time scale 
$\tau$ is defined as follows: For a given operator $\hat{O}$, 
one 
first defines the maximum value of $\tau$ which satisfies 
(3.10) for each $\hat{\rho}$. Based on the set of those values 
$\{\tau(\rho) \}$, one may define the intrinsic time scale
associated with $\hat{O}$ by the minimum of $\tau(\rho)$,
namely, $\tau=\min_{\rho}\{\tau(\rho) \}$. In practice,
however, the time dependence of an observable always appears
in the combination $Tr\hat{\rho}(t)\hat{O}$ in our treatment.
One may thus use the time scale given by the maximum of $\tau$
which satisfies (3.10) for a given physical system 
 specified by $\hat{\rho}(t)$.
\bibitem{f8}
One may avoid the recurrence theorem by appealing to the 
thermodynamic limit with $V\rightarrow large$ and  the 
particle density $N/V$ kept finite, but at the same time one
looses the 
dynamical information of the system. We know very little about 
the dynamical properties of a many particle system, and the 
reccurence theorem provides precious information.
\bibitem{f9}
For a particle
number $N=10^{23}$ at room temperature, $\Delta E\sim kT\times
\sqrt{N}\sim 10^{10}eV$. If one chooses $\Delta t_{c}\sim
10^{-1}sec$, for example, 
$2\pi\hbar/\Delta t_{c}\sim 10^{-13}eV$.
\bibitem{f10}
A similar consideration is applicable to the 
analysis of a time dependent Hamiltonian if one analyzes the 
behavior of the system after switching off the time dependent 
perturbation.
\bibitem{f11}
This picture suggests that the bulk of 
entropy generation takes place within the time scale of 
relaxation
measured by macroscopic observables.
\bibitem{f12}
The essence of the inequality in ref. 39 
is stated in our framework as follows: One may denote the 
density matrix for a (micro-)canonical ensemble (3.6) by 
$\hat{\rho}_{0}$ and the corresponding average energy by $E_{0}$.
After an arbitrary small unitary transformation, we have 
$\hat{\rho}=U\hat{\rho}_{0}U^{\dagger}$ and 
$E=\sum_{n}E^{\prime}_{n}\langle n|\hat{\rho}|n\rangle$. Our 
inequality
(3.8) indicates $S=-k\sum_{n}\langle n|\hat{\rho}|n\rangle
\ln\langle n|\hat{\rho}|n\rangle\geq 
H=-k Tr\hat{\rho}\ln\hat{\rho}
=-kTr\hat{\rho}_{0}\ln\hat{\rho}_{0}=S(E_{0})$. The maximum of 
$S$ with fixed $E$, which is the physical entropy if  
relaxation takes place, thus satisfies $S(E)\geq S(E_{0})$. The 
condition $dS(E_{0})/dE_{0}=k\beta(E_{0})>0$ then implies
$E\geq E_{0}$ for an arbitrary small unitary transformation
$U$, which includes time development generated by a time-dependent
Hamiltonian.${}^{39)}$ Our analysis in this paper is 
complementary to that of ref. 39 where no analysis of relaxation 
after an external perturbation is given.
\end{thebibliography}
\end{document}